\documentclass[fleqn,10pt,conference,a4paper]{IEEEtran}
\usepackage[normalem]{ulem}
\usepackage{amsmath,amssymb,float} 
\usepackage{graphicx}
\usepackage{cite}
\usepackage{flushend}[keeplastbox] 
\usepackage{color}                 
\usepackage{xcolor}
\usepackage{fancyhdr}
\usepackage{url}
\usepackage{subcaption} 
\usepackage{comment}

\fancypagestyle{firstpage}
{
    \fancyhf{}
    \fancyhead[C]{The paper has been presented at the 2025 35th International Conference Radioelektronika (RADIOELEKTRONIKA),\\ Hnanice, Czech Republic, May 12-14, 2025, \url{https://doi.org/10.1109/RADIOELEKTRONIKA65656.2025.11008377}}
    \fancyfoot[C]{This research was funded in part by the National Science Center (NCN), Poland, grant no. 2021/43/I/ST7/03294 (MubaMilWave). For this purpose of Open Access, the author has applied a CC-BY public copyright license to any Author Accepted Manuscript (AAM) version arising from this submission.}
}

\begin{document}

\title{Framework for Indoor Wireless Propagation Modeling through Wireless Insite\textsuperscript{\textregistered}}
\author
{
\IEEEauthorblockN{Arao~Zau~Macaia\IEEEauthorrefmark{1},
Niraj~Narayan\IEEEauthorrefmark{1},
Rajeev~Shukla\IEEEauthorrefmark{1},
Aniruddha~Chandra\IEEEauthorrefmark{1},
Ondřej Zelený\IEEEauthorrefmark{2},\\
Radek~Zavorka\IEEEauthorrefmark{2},
Jiri~Blumenstein\IEEEauthorrefmark{2},
Ale{\v{s}}~Proke{\v{s}}\IEEEauthorrefmark{2},
Jarosław~Wojtuń\IEEEauthorrefmark{3},
Jan~M.~Kelner\IEEEauthorrefmark{3},
Cezary~Ziolkowski\IEEEauthorrefmark{3}
}
\IEEEauthorblockA{
\IEEEauthorrefmark{1}ECE Department, NIT Durgapur, 713209 Durgapur, India (aniruddha.chandra@ieee.org)
}
\IEEEauthorblockA{
\IEEEauthorrefmark{2}Department of Radio Electronics, Brno University of Technology, 61600 Brno, Czech Republic
}
\IEEEauthorblockA{
\IEEEauthorrefmark{3}Institute of Communications Systems, Faculty of Electronics, Military University of Technology, Warsaw, Poland }
}
\maketitle
\thispagestyle{firstpage}

\begin{abstract}
Multipaths, reflections, diffractions, and material interactions complicate indoor wireless propagation modelling. More than $80\%$ of wireless data is consumed indoors; hence, planning successful deployments and maximizing network performance depends on accurate propagation modelling of indoor environments. This work explains a complete framework for indoor wireless propagation modelling via ray tracing simulation in a step-by-step manner. The ray tracing simulations are conducted with Wireless Insite®, a proprietary electromagnetic propagation software, whereas SketchUp® is used at the input side for layout construction from the field measurements, and MATLAB® is used at the output side for portraying channel model parameters such as power delay profile (PDP). A whole floor of the authors' department is modelled, and different transmitter-receiver locations were tested for possible use cases such as coverage hole prediction.
\end {abstract}


\section{Introduction}\label{sec:intro}

\subsection{What is Ray Tracing?}
In wireless channel modelling, ray tracing (RT) is a ray optics-inspired technique used to represent the propagation of radio waves through a given indoor or outdoor environment \cite{kanhere2024calibration,aghaei2024vehicular}. The way waves move from the transmitter to the receiver is examined, taking into account their interactions with objects like trees and buildings, which can cause the waves to disperse, bounce, or bend \cite{zhang2024deterministic, Laurenson1993}. Using this approach, a three-dimensional (3D) map of the surroundings is produced by tracking the waves' courses and illustrating how they alter as they strike various surfaces \cite{chang1998}. Ray tracing helps predict signal strength, interference, and coverage in complicated regions, such as built-up urban areas or highly cluttered industrial environments, where traditional models may not perform well \cite{Esposti2021}. Radio waves behave like light at higher frequency ranges, making the ray tracing-based framework more relevant in millimetre wave (mmWave) and sub-terahertz (sub-THz) bands.

\subsection{Ray Tracing for Indoor Environments}
Indoor modelling is used to understand how wireless (like WiFi or phone) signals behave inside buildings. Since Global Positioning System (GPS) signals and satellite images do not work well indoors, measuring and predicting signal strength and coverage is harder. This is where ray tracing comes in. Ray tracing helps by simulating how radio waves bounce off walls, furniture, and other objects inside the building, giving us an idea of how strong the signal will be in different areas \cite{Seidel1992}. Ray tracing is also helpful in checking the accuracy of models. After taking real measurements inside the building, engineers can compare them to the ray tracing results to see if the model is correct\cite{Laurenson1993}. This helps improve the models so they reflect the real world more accurately.
Additionally, ray tracing is helpful for interpolation and extrapolation. Engineers can use measurements taken at a few points inside the building and then predict the signal strength in other areas without needing to measure everywhere\cite{Verdecia2023}. It helps save time and effort while providing accurate signal coverage predictions.

\begin{figure}[t]
    \centering
    \includegraphics[width=0.48\textwidth]{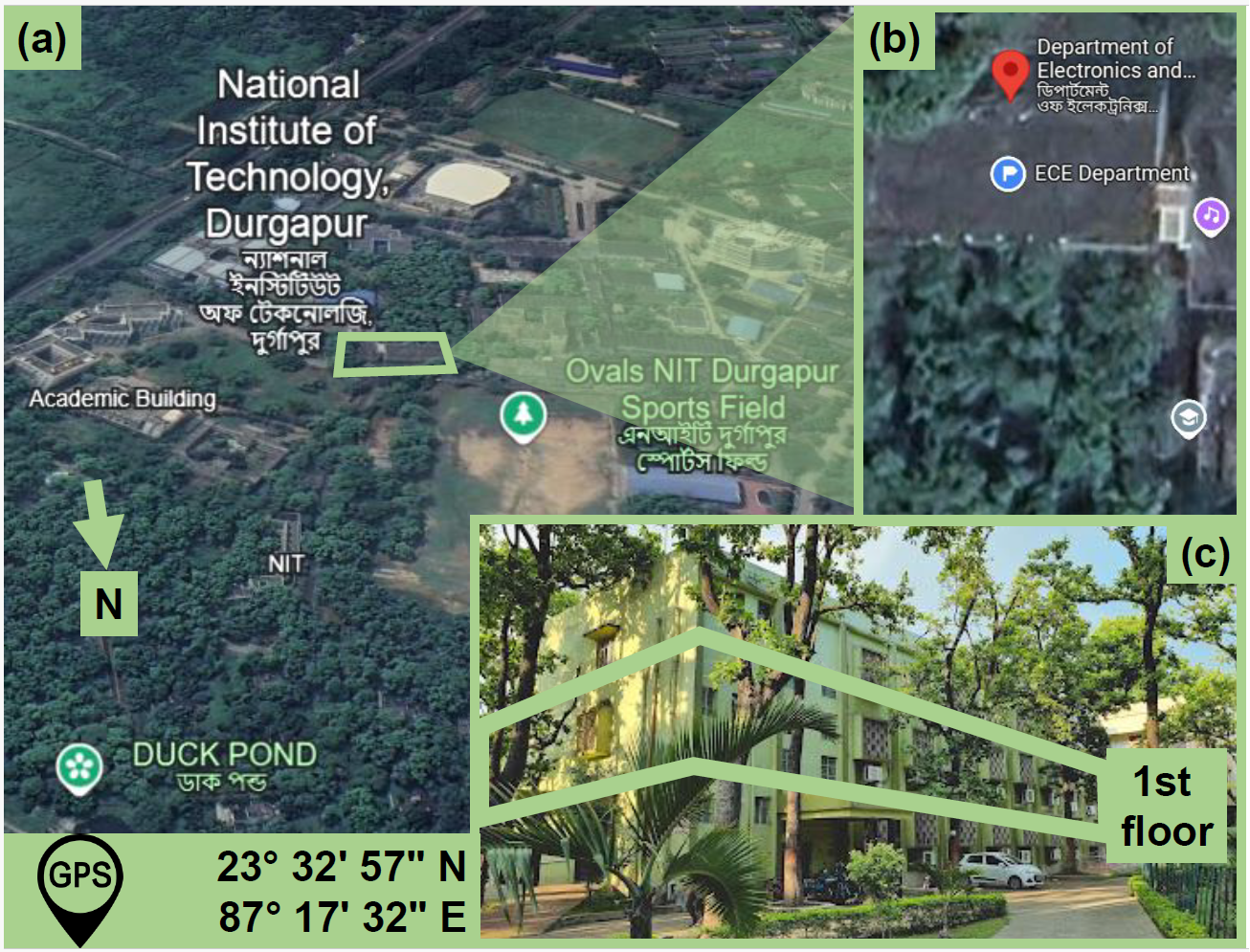}
    \caption{Satellite images and building photograph of the site [clockwise from left]: (a) Aerial view of National Institute of Technology (NIT) Durgapur, (b) Top view of the Electronics and Communication Engineering (ECE) department, and (c) Photograph of the ECE department building taken from south-west direction. [image courtesy: Google Earth, Google Maps]}
    \label{Fig.1}
\end{figure}

\subsection{Literature Survey}
Ray tracing is widely applied in indoor wireless propagation modelling to simulate radio wave behaviour and predict key parameters like signal strength, delay spread, and coverage holes. Authors in \cite{zhu2022learning} presented an end-to-end learning architecture that combines significance sampling and differentiable Monte Carlo ray tracing. The method enables high-fidelity material editing and complex item insertion by simultaneously recovering geometry, spatially varying illumination, and photorealistic materials from a single image.
Using detailed ray-tracing simulations, the article \cite{Topal_RT_2022} describes a 28 GHz indoor dense space channel while considering settings such as train waggons and aeroplane cabins. Comparing bit error rate performance to current models, it examines route loss, shadow fading, and other factors. The authors conducted a comprehensive indoor environment characterization \cite{OBEIDAT2020RT} and provided a detailed discussion on various channel parameters and channel models, the influence of building material on channel propagation, and some case studies. 
\begin{figure}[t]
    \centering
    \includegraphics[width=0.48\textwidth]{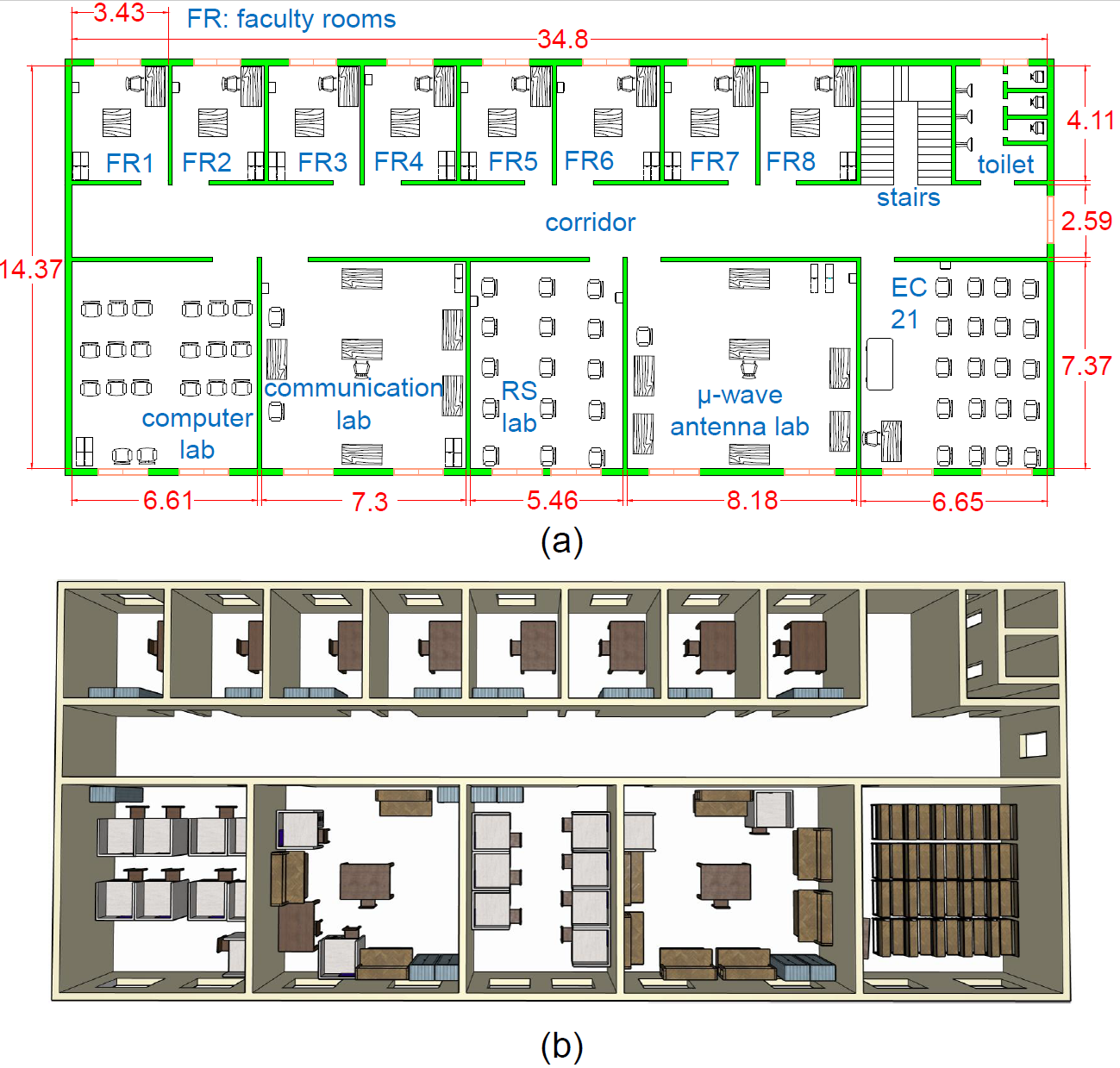}
    \caption{Layout of the 1st floor (a) Top view designed in AutoCAD and (b) Three-dimensional (3D) view designed in SketchUp. The corridor is stretched east-west, and faculty rooms are south-facing. Dimensions are in m.}
    \label{Fig.2}
\end{figure}

\begin{figure}[h]
    \centering
    \includegraphics[width=0.48\textwidth]{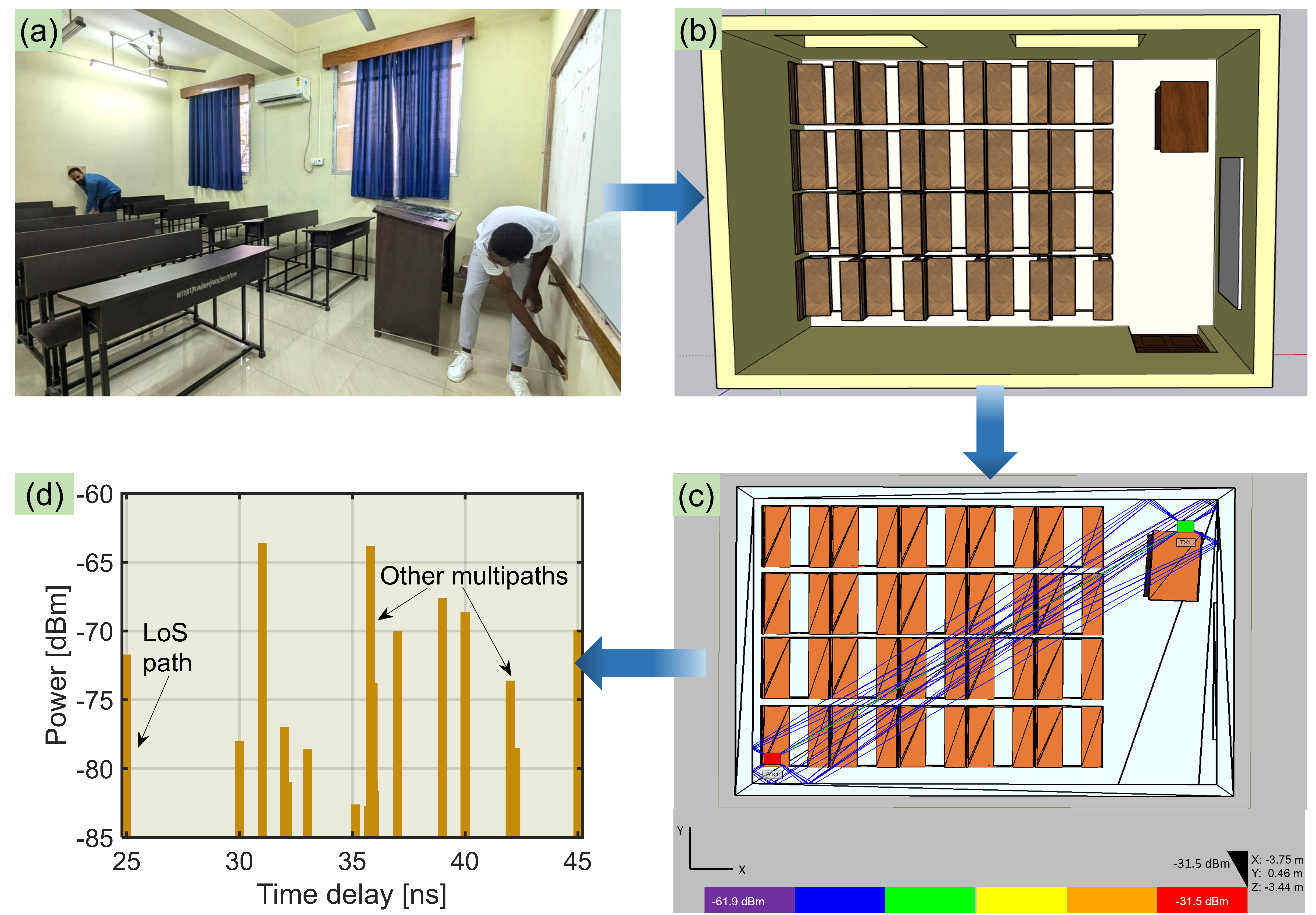}
    \caption{Four stages of the workflow [clockwise from top-left]: (a) Dimensional measurements are being taken in the classroom EC 21, (b) 3D layout is obtained in SketchUp after measurements are fed, (c) Indoor radio propagation environment layout is constructed with Wireless Insite, and (d) Channel model parameters are viewed through MATLAB.}
    \label{Fig.3}
\end{figure}

Using indoor geolocation algorithms, authors in \cite{Clark2023RTInsite} first simulated an indoor scenario using Wireless InSite and evaluated the geolocation performances. Potentials and limitations of channel characterization characterization $5$ GHz using the RT tool of Wireless InSite were introduced in \cite{Akram2019IndoorInsite} for an indoor propagation scenario. The authors investigated the effect of distance, frequency and obstacles in line-of-sight (LOS) and non-line-of-sight (NLOS) scenarios. On the other hand, \cite{Abed2022CoverageInsite} applies Wireless Insite® to optimize the coverage of the WiFi network in a large office building. The authors use ray tracing to simulate different access point configurations and determine the most efficient setup. The paper provides a case study of how Wireless Insite® can be used for real-world network optimization, focusing on range optimization, hole detection and network planning in commercial spaces.

The organization of the rest of the paper is as follows. Section II describes a step-by-step guide on using the RT software for indoor wireless propagation modelling. A full description of the example site model, the four main stages of the workflow, and verification methods are presented. Section III presents the methodology and results for a typical use case, coverage hole prediction. The paper finally concludes in Section IV with an indication of the future scope of the framework presented.  
\begin{figure*}
    \centering
    \includegraphics[scale=0.13]{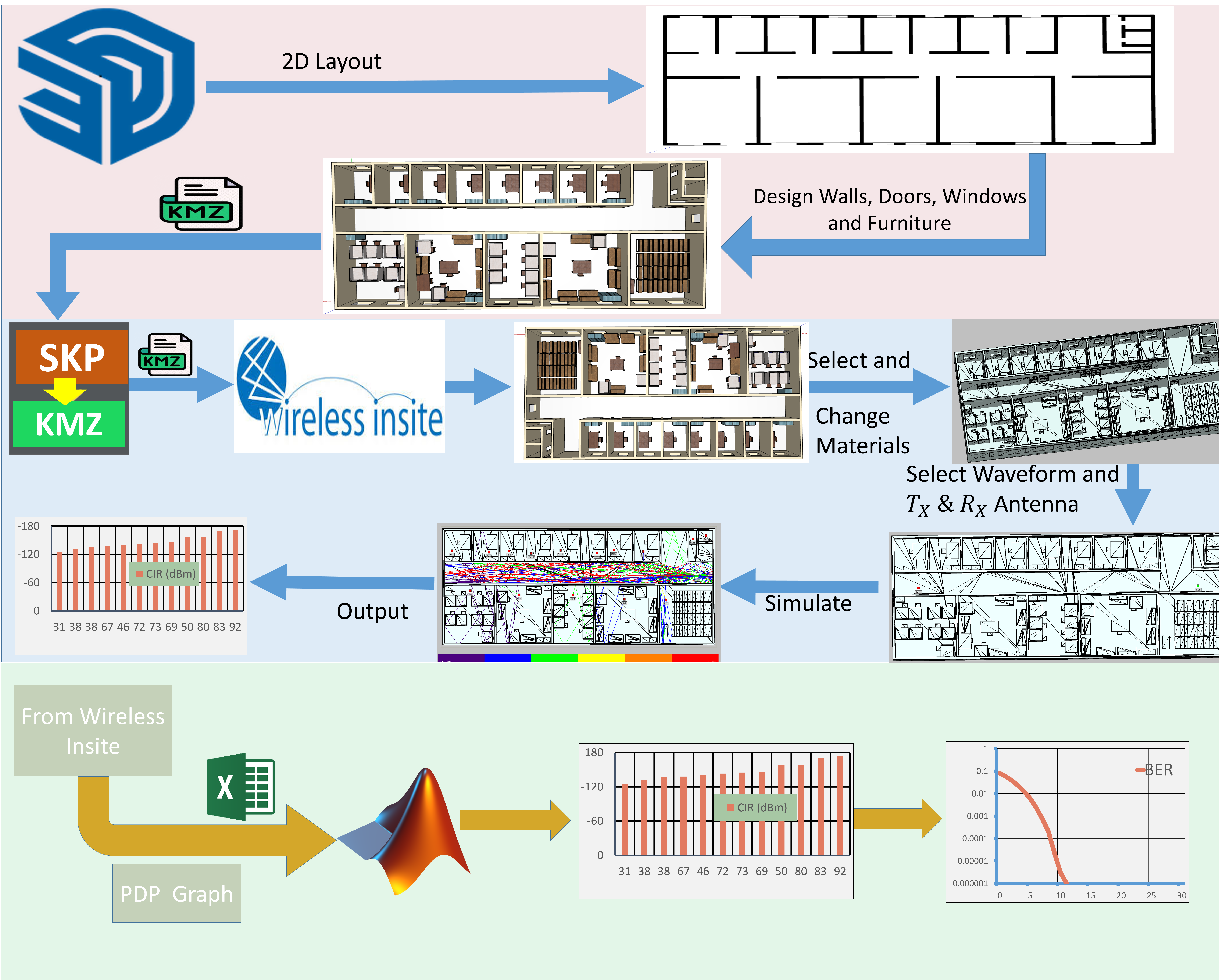}
    \caption{Detailed workflow in a step-by-step manner - the top portion shows SkechUp flow, the middle layer shows the main Wireless Insite flow, and the bottom layer shows the MATLAB flow.}
    \label{Fig.4}
\end{figure*}

\begin{figure*}
    \centering
    \includegraphics[scale = 0.53]{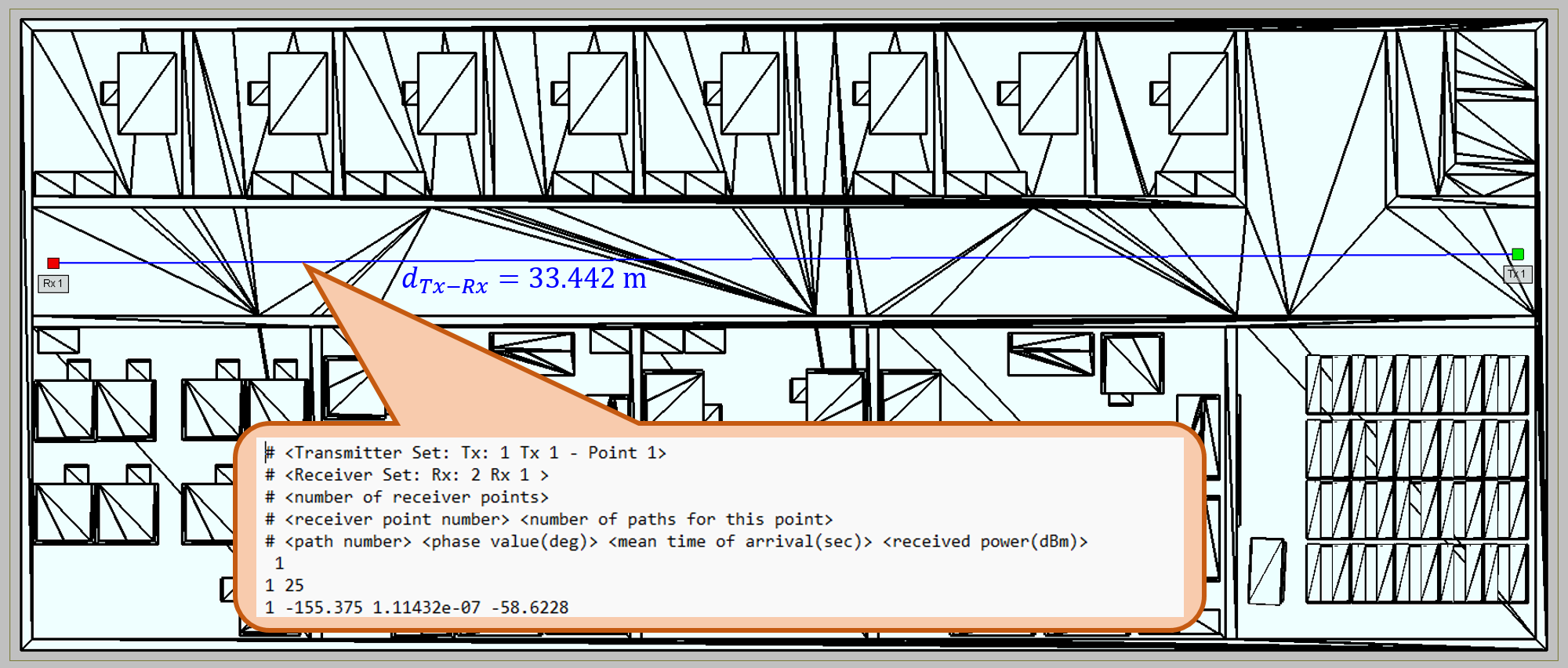}
    \caption{Verification of the constructed radio propagation environment. The distance between $T_X$ and $R_X$ divided by the propagation speed should match the delay of the first-arriving LOS path delay.}
    \label{Fig.5}
\end{figure*}

\begin{figure*} []
    \centering
    \subfloat[\centering Wooden Furnitures]
    {\includegraphics[scale=0.09]{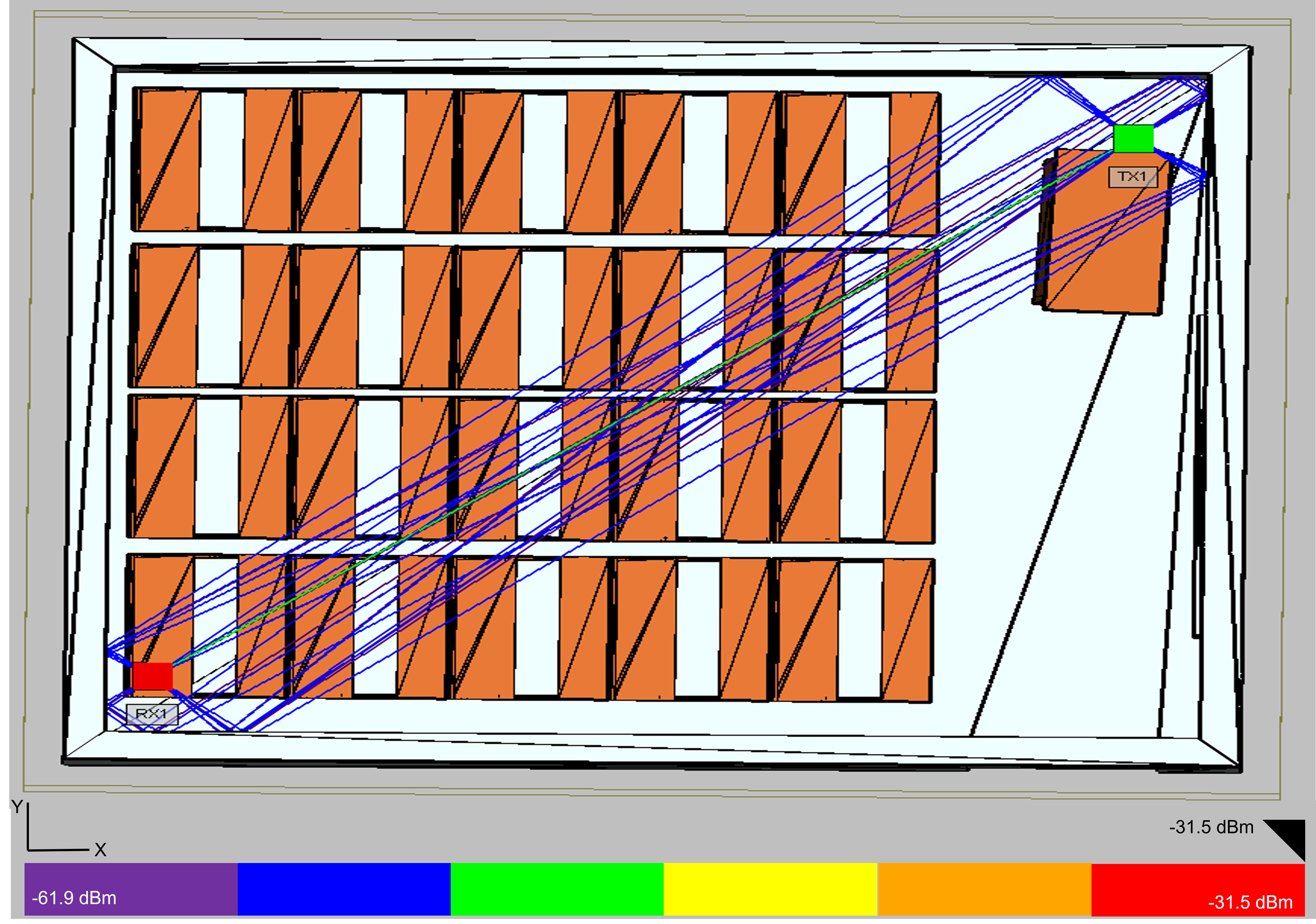}}
     \subfloat[\centering Metallic Furnitures]{\includegraphics[scale=0.09]{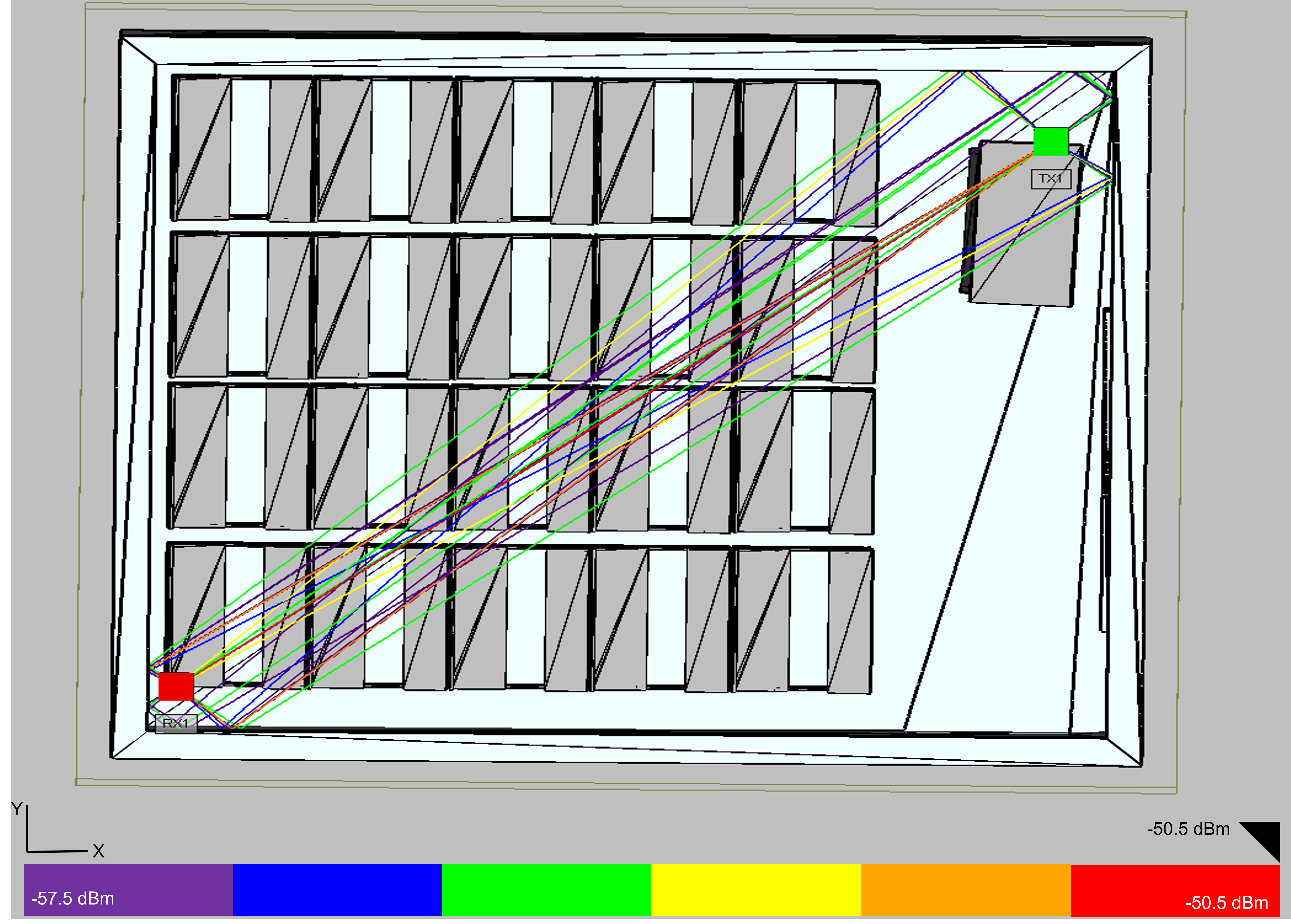}}
    \caption{Ray tracing results for the same environment with two different types of material: [left] With all wooden furniture, and [right] With metal coating on furniture.}
    \label{Fig.7}%
\end{figure*}

\section{Workflow with Wireless Insite}

\subsection{Site Description}
The example site that is modelled with the RT framework is the 1st floor of the Electronics and Communication Engineering (ECE) department of the National Institute of Technology (NIT) Durgapur, an institute of national importance situated in the eastern side of India in the state of West Bengal. The plus code of the site is G7XR+MX Durgapur, West Bengal. Fig. \ref{Fig.1} gives a pictorial description of the exterior of the modelled site.

The first floor's interior, as shown in Fig. \ref{Fig.2}, consists of eight faculty rooms and five laboratories arranged on the right and left sides of a corridor, access to which is through a staircase. The height of the floor from the ground to the ceiling is $3.44$m. The length of the floor is $34.8$m, and the width is $14.37$m. The laboratories on the left side have the following dimensions: computer lab $6.76$m$\times7.37$m, communication lab $7.3$m$\times7.37$m, RS lab $5.45$m$\times7.37$m, microwave and antenna lab $8.17$m$\times7.37$m and EC 21 classroom $6.64$m$\times7.37$m. Faculty rooms are all of equal dimensions $3.43$m$\times4.1$m.

\subsection{Stages of the Workflow}
Fig. \ref{Fig.3} summarizes the who summarizes as a four-stage process. In the first stage, we took measurements with a meter tape and pocket laser distance measurer (model no: BOSCH GLM-40). In the second stage, these measurements were used to design a 3D model of the building floor using SketchUp. The third stage involves exporting the SketchUp file to Wireless InSite, where the RT simulation is performed. The output of the RT tool in the form of a power delay profile (PDP) was constructed with MATLAB in the fourth and final stage.

A detailed version of the workflow is presented in Fig. \ref{Fig.4}. To create the 3D model of our department's first floor using SketchUp, we began by drawing the 2D layout based on the measurements we had taken from the actual building. We used tools such as the Line Tool, Rectangle Tool, and Tape Measure Tool to accurately outline the structure, including the walls and room divisions. After completing the 2D layout, we raised the walls using the Push/Pull (Extrude) Tool to convert the design into a 3D model. Openings for doors and windows were created using a combination of the Line Tool, Tape Measure Tool, and Push/Pull Tool. We then proceeded to add doors, windows, and furniture, either by manually modelling them or importing components from the SketchUp 3D Warehouse, assigning appropriate materials and textures to each. Once the structural and interior elements were in place, we painted the interior and exterior walls using the Paint Bucket Tool, selecting suitable materials and colours from the materials library to make the model as realistic as possible. Finally, we saved the file in \texttt{.skp} format and exported it as a \texttt{.kmz} file to be imported into Wireless InSite for further simulation and analysis. 

In Wireless Insite, after defining the structural elements, we proceed to waveform selection and choose the appropriate antenna type, such as an isotropic or horn antenna. Transmitter ($T_X$) and receiver ($R_X$) points are placed strategically in the model. This is followed by a selection of output parameters like propagation path, path loss, complex impulse response (CIR), and delay spread. Finally, we run the simulation, and the resulting output is obtained. MATLAB may be used to construct PDP from the CIR values given by the Wireless Insite simulation. Subsequently, a tapped-delay-line channel model from the PDP can be built and used for numerous purposes, including bit-error-rate (BER) simulation.

\subsection{Verification}
The verification of the output is a necessary step where the first reading of the CIR, i.e. the delay of the first-arriving LOS path, is used to verify whether the scaling of the parameters is correct. An example of such a verification step is shown in Fig. \ref{Fig.5}, where two isotropic antennas are positioned as $T_X$ and $R_X$ in the corridor. Measurement tool in Wireless Insite gives that the distance between $T_X$ and $R_X$ is, $d_{T_X-R_X} = 33.44$m. The simulation output for such a setting is shown in the textbox of Fig. \ref{Fig.5}, where the last line of the text output gives a mean time of arrival, $ToA=111.43$ nanoseconds (ns). If we multiply $ToA$ with the speed of light ($c=3\times 10^8$m/s), we get the estimated path length as, $\hat{d}_{T_X-R_X} = ToA \times c\thickapprox 33.43$m. If we obtain $\hat{d} \thickapprox d$ for a series of such verification tests, which might be necessary if the constructed environment is non-homogeneous, we may use the results from RT simulation with confidence.

\begin{figure}[ht]
    \centering
    \includegraphics[width=0.44\textwidth]{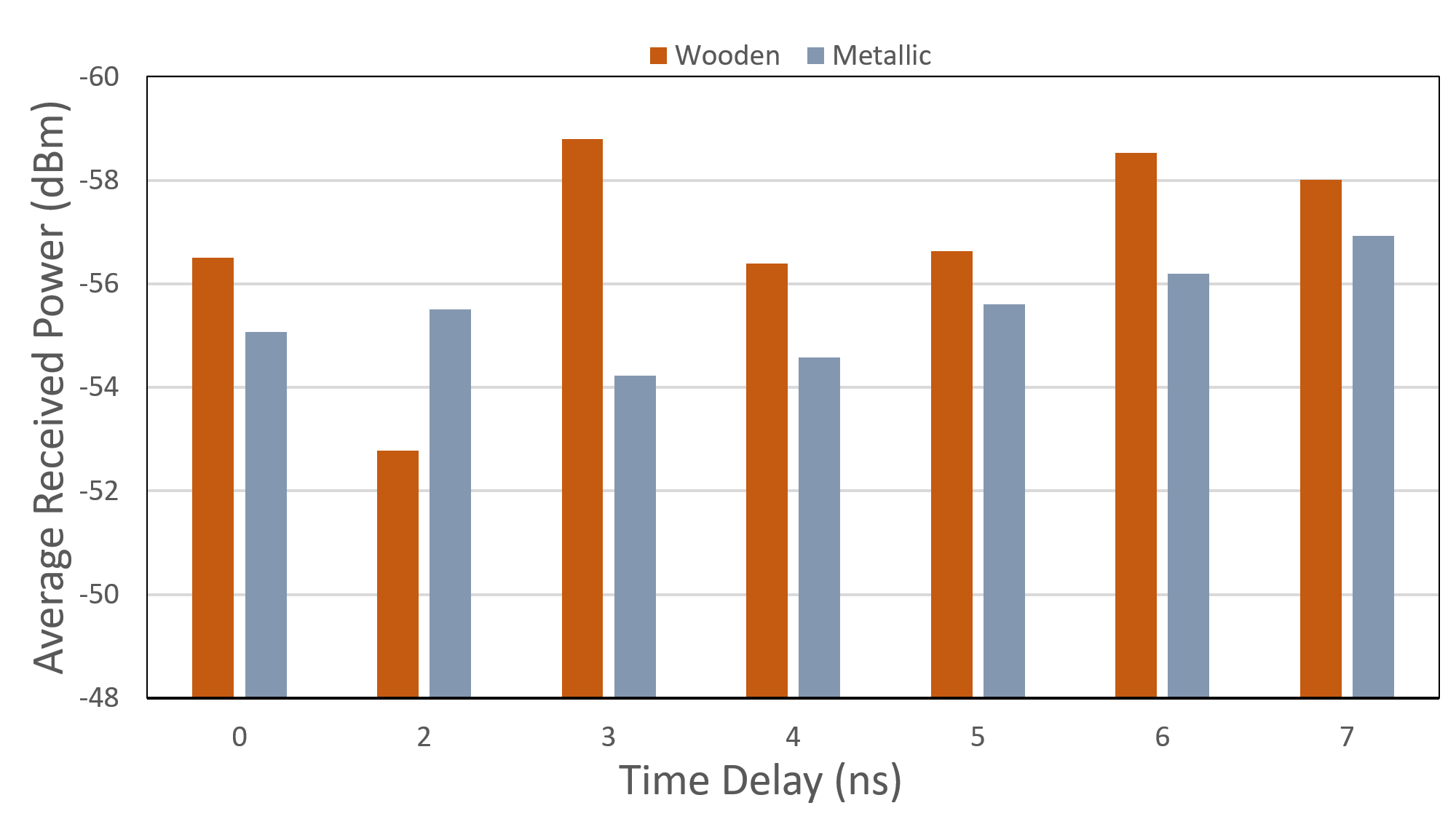}
    \caption{Effect of change in materials in the received signal strength across different propagation paths.}
    \label{Fig.8}
\end{figure}

A second set of verification is done next for material properties. We have run the simulation for the same classroom (EC 21) with wooden furniture first, and then made the furniture surfaces metallic. The results are shown in Fig. \ref{Fig.7}, where the left figure shows rays in dominantly blue, representing a lower signal strength due to absorption by wooden material, whereas the figure on the right gives many dominant paths (shown with yellow/ green) apart from the LoS path due to strong reflection from metal. The same is visible in Fig. \ref{Fig.8}, where a one-to-one comparison of traced rays is made.

\begin{figure*}[th]
    \centering
    \includegraphics[width=\textwidth]{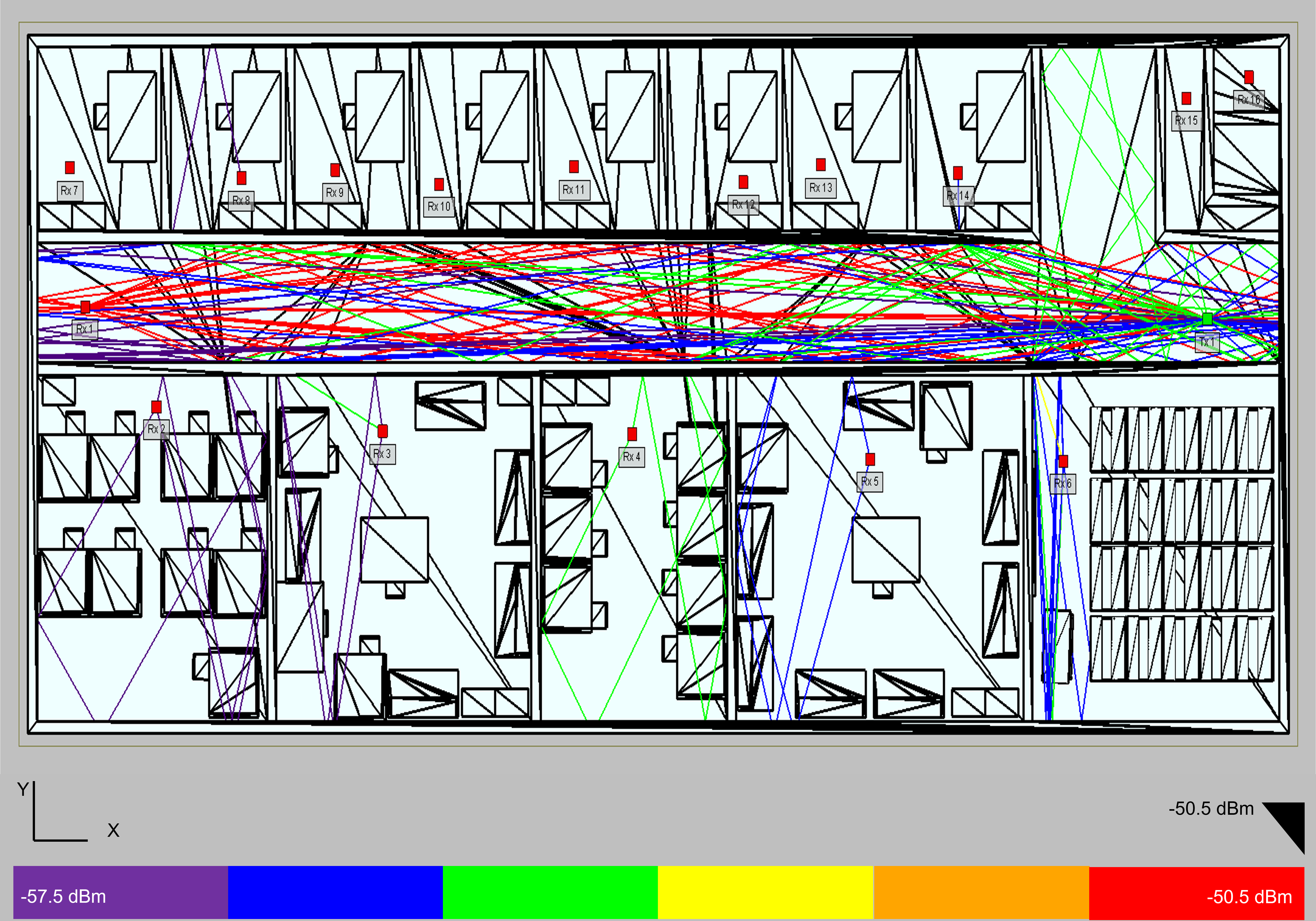}
    \caption{Simulation results from Wireless Insite showing the major propagation paths for a given $T_X$ and 16 possible $R_X$s placed randomly throughout the floor.}
    \label{Fig.6}
\end{figure*}

\begin{figure}[!t]
    \centering
    \includegraphics[scale=0.2,trim = 11cm 0cm 13cm 2cm]{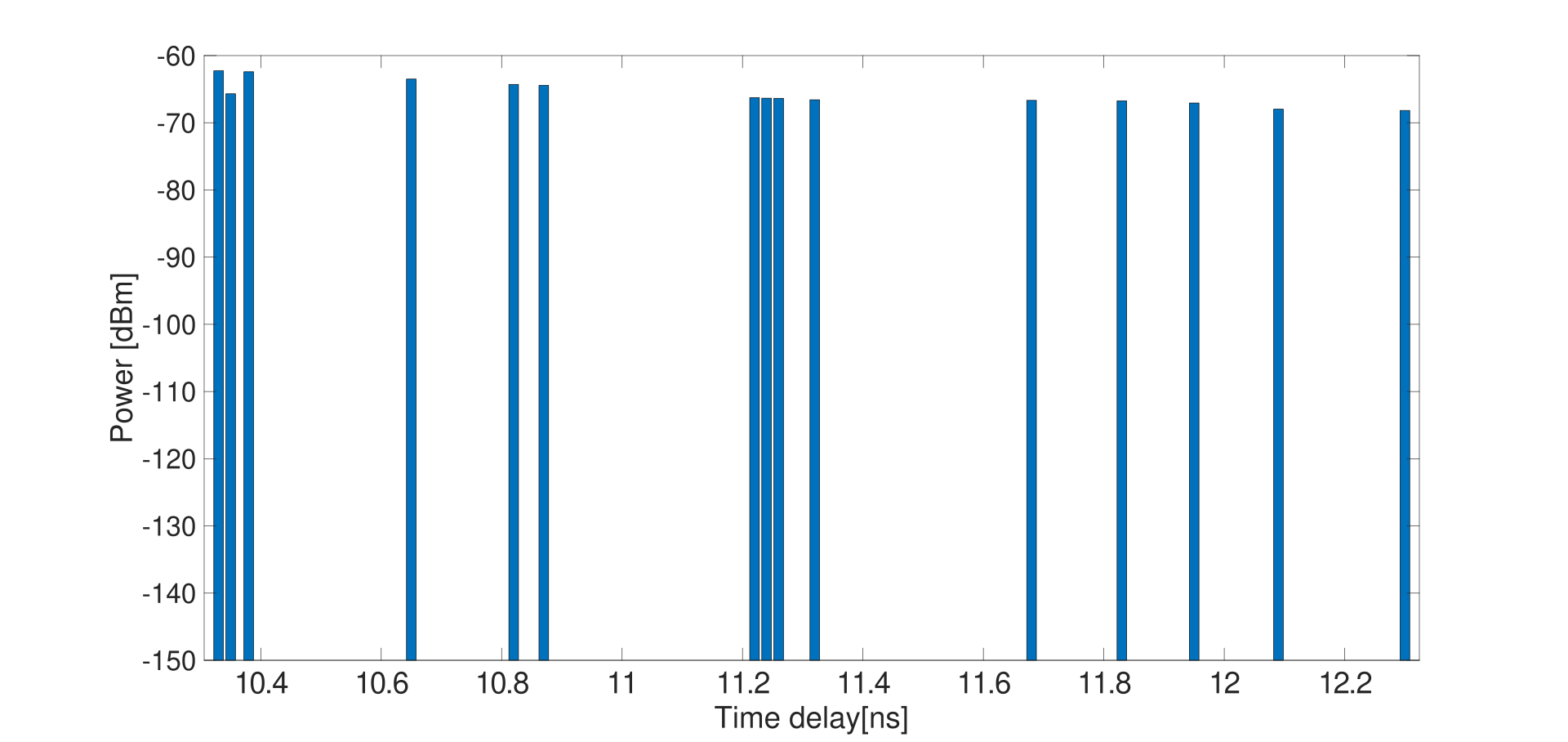}
    \caption{PDP for channel between $T_{X1}$ and $R_{X1}$.}
    \label{Fig.9}
\end{figure}
\begin{figure}[!t]
    \centering
    \includegraphics[scale=0.2,trim = 11cm 0cm 13cm 2cm]{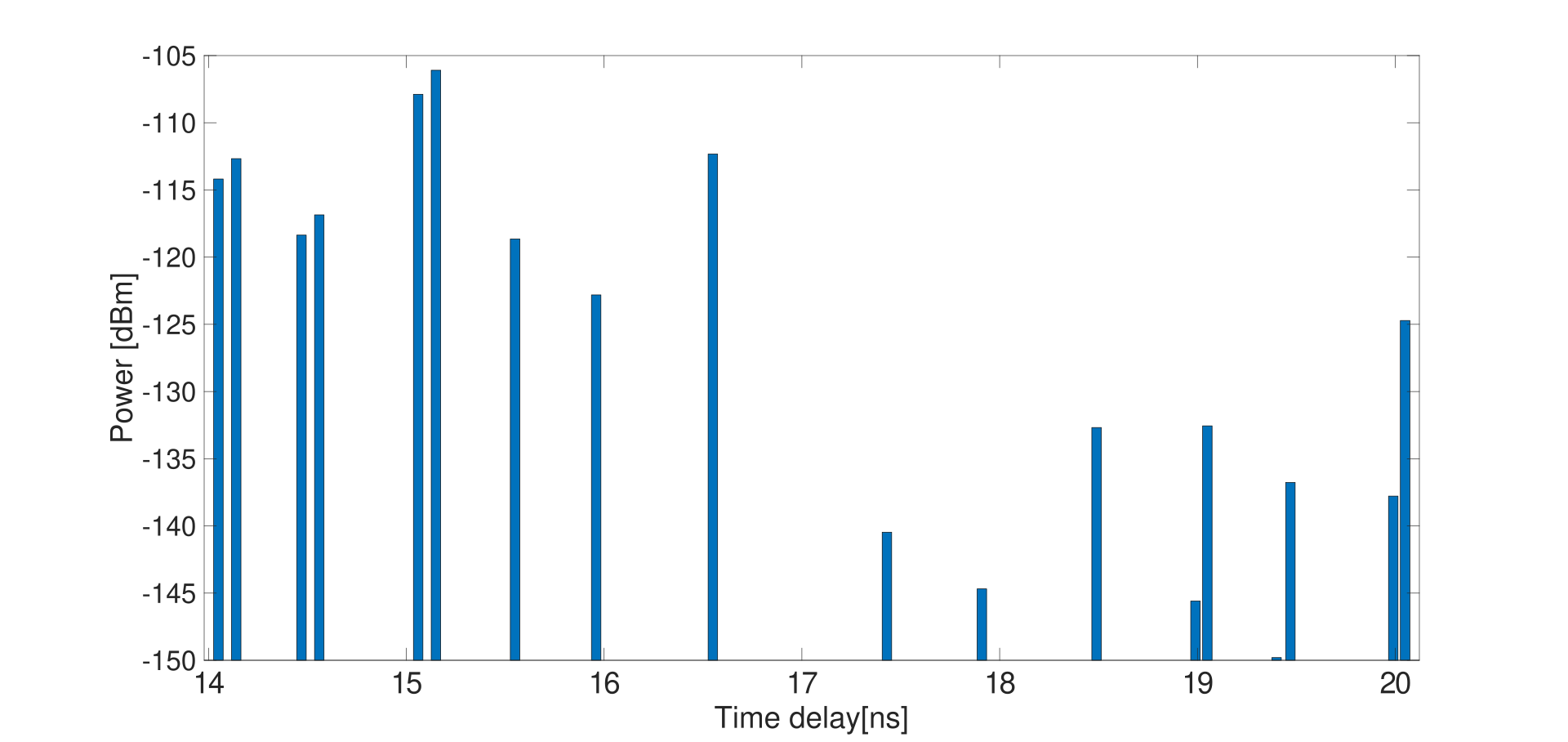}
    \caption{PDP for channel between $T_{X1}$ and $R_{X2}$.}
    \label{Fig.10}
\end{figure}

\section{Use Case of the WorkFlow}
The workflow presented in the previous subsection can be utilized for performance predictions related to wireless signal propagation. As a typical example, we will explain the idea of coverage hole prediction in this section. 

We have set up an isotropic antenna as a single $T_X$ and placed it at the eastern side of the corridor, as shown in Fig. \ref{Fig.6}. We have also set up 16 receivers, $R_{X1}, \cdots, R_{X16}$, inside various rooms and laboratories. The simulation parameters are as follows: signal - sinusoidal, centre frequency - $1$ GHz, bandwidth - $10$ kHz, maximum number of rays - $25$, antenna height - $2$m. A full simulation gives the dominant paths as shown in Fig. \ref{Fig.6}. We found many coverage holes, especially all the faculty rooms are devoid of a strong signal. Thus, the initial $T_X$ position is not optimum, and wireless planning on the floor requires further iterations.

To illustrate the effect further, we provide the PDP for two receivers, $R_{X1}$ and $R_{X2}$, in Figs. \ref{Fig.9} and \ref{Fig.10}, respectively. For a receiver in the corridor ($R_{X1}$), the average signal strength is about 40dB above the signal level of the receiver in a laboratory area  ($R_{X2}$).

\section{Future Work}
In future, we are planning to perform a WiFi access point (AP) location optimization using the presented framework, where we will divide the whole public area into $1$m$\times1$m grid and check the $T_X$ position with maximum AP coverage. This simulation will then be verified using actual physical measurements. 

Other interesting extensions include validating measurement-based models through RT models when measurement results for the same set of $T_X$ and $R_X$ locations are available. Further, RT simulation through other software (e.g. Sionna, MATLAB RF propagation toolbox, NIST Q-D tool, RANPLAN) can also be cross-validated.




\section*{Acknowledgement}
This work was developed within a framework of the research grants: project no. 23-04304L sponsored by the Czech Science Foundation, MubaMilWave no. 2021/43/I/ST7/03294 funded by the National Science Centre, Poland, under the OPUS call in the Weave programme, grant no. UGB/22-748/2024/WAT sponsored by the Military University of Technology, and chips-to-startup (C2S) program no. EE-9/2/2021-R\&D-E sponsored by MeitY, Government of India.

\bibliographystyle{./IEEEtran}    
\bibliography{./References}


\end{document}